\documentstyle[12pt]{article}
\begin{document}

\vspace*{1cm}
\begin{center}
Invited paper for the ``International Workshop on Antimatter Gravity
and Antihydrogen Atom Spectroscopy'', Sepino (IS), Italy, May 1996. \\
To be published in Hyperfine Interactions.
\end{center}
%Invited paper for the `` Internationale Workshop on Antinua ther gravity
%and Antivy drogen atom Spectroscopy'',sepino (is), Italy, May 1996. To
%be published in Hyper fine Interactions.

\begin{center}
{\huge \bf DOES ANTIMATTER EMIT \\[0.5cm] A NEW LIGHT ?}
\end{center}

\begin{center}
{\large \bf Ruggero Maria Santilli\footnote{Permanent address:
Institute for Basic Research, P.O.Box 1577,Palm
Harbor, FL 34682,U.S.A., \\
\hspace*{0.7cm}ibrrms@pinet.aip.org}}
\end{center}

\begin{center}
{\large Instituto per la Ricerca di Base \\ Molise, Italy}
\end{center}

\begin{abstract}
Contemporary theories of antimatter have a number of insufficiencies
which stimulated the recent construction of the new {\it isodual theory}
based on a certain anti-isomorphic map of all (classical and quantum)
formulations of matter called {\it isoduality}. In this note we show that the
isodual theory predicts that antimatter emits a new light, called
{\it isodual light}, which can be distinguished from the ordinary light
emitted by matter via gravitational interactions (only). In
particular, the isodual theory predicts that all stable antiparticles such
as the isodual photon, the positron and the antiproton experience
antigravity in the field of matter (defined as the reversal of the sign
of the curvature tensor). The antihydrogen atom is therefore
predicted to: experience antigravity in the field of Earth; emit the
isodual photon; and have the same spectroscopy of the hydrogen atom,
although subjected to an anti-isomorphic isodual map. In this note we
also show that the isodual theory predicts that bound states of
elementary particles and antiparticles (such as the positronium)
experience ordinary gravitation in both fields of matter and
antimatter, thus bypassing known objections against antigravity. A
number of intriguing and fundamental, open theoretical and
experimental problems of ``the new physics of antimatter'' are pointed
out.
\end{abstract}

\section{INTRODUCTION}
Since the time of Dirac's prediction of antiparticles and their
detection by Anderson (see~\cite{ref1} for historical accounts ), the theory
of antimatter has been essentially developed at the level of {\it second
quantization}.

 This occurrence has created an unbalance between the theories of matter
and antimatter at the {\it classical} and {\it first quantization} levels, as well
as a number of shortcomings, such as the inability for the classical
theory of antimatter to have a quantized formulation which is the
correct charge (or PTC) conjugate of that of matter.

In an attempt  to initiate the scientific process toward the future
resolution of the above problematic aspects, this author proposed in
1985~\cite{ref2} a new anti-isomorphic image of conventional mathematics
charactered by the map of the conventional unit
\begin{equation}
+1\rightarrow 1^d=-1^\dagger=-1,
\label{eq:one-one} \end{equation}
called for certain technical reasons {\it isodual map}, or {\it isoduality}.

It should be noted that the change of the basic unit implies a simple,
yet unique and non-trivial change of the totality of conventional
mathematics, including: numbers and angles; functions and transforms;
vector and metric spaces; algebras and geometries; etc.

In 1991 this author~\cite{ref3} showed that the above isodual mathematics,
since it is an anti-isomorphic image of the mathematics of matter,
provides a novel {\it classical} representation of antimatter.

The proof that isoduality on a Hilbert space is equivalent to charge
conjugation first appeared in paper~\cite{ref4} of 1994. A comprehensive
operator treatment subsequently appeared in monographs~\cite{ref5}.

The prediction that isoduality implies {\it antigravity} (defined as the
reversal of the sign of the curvature tensor) for massive antipaticles
in the field of matter was submitted in paper\cite{ref6}, which also included
the proposal for its experimental verification via the use of a low
energy~(eV) positron beam in horizontal flight in a suitable vacuum
tube. The latter experimental proposal was subsequently studied by
Mills~\cite{ref7}.

This note is devoted to a study of the {\it spectroscopy of antimatter} via
the isodual characterization of the light emitted by the antihydrogen
atom~\cite{ref8}. In particular, we show that {\it isoduality predicts that
antimatter emits a new light here called isodual light} which can be
solely differentiated from the conventional light via gravitational
interactions.

In the events additional theoretical and experimental studies confirm
the above hypothesis, isoduality would therefore permit the future
experimental measures whether far away galaxies and quasars are
made-up of matter or of antimatter.

A more comprehensive analysis is presented in memoir~\cite{ref9}, which also
includes the study of isodual theories of antimatter at the more
general {\it isotopic} and {\it isogravitational levels}.
Ref.\cite{ref9} also shows that
know objections against antigravity are {\it inapplicable} to (and not
``violated'' by) the isodual theory of antimatter, because the former
are (tacitly) constructed on conventional mathematics, while the latter is
 formulated on a novel mathematics based on a new negative unit.

After reviewing the mathematical foundations and the main predictions
of the isodual theory, in this note we identify a number of rather
fundamental, open, theoretical and experimental problems of the
emerging ``new physics of antimatter''.

\section{ISODUAL MATHEMATICS}

Our fundamental assumption is that of antimatter is that
it is represented by the new {\it isodual mathematics} which is the
anti-isomorphic image of conventional mathematics under map~(\ref{eq:one-one}).

Since the latter mathematics is still vastly unknown (in both
mathematical and physical circles), it appears recommendable to
outline in this section the main notions in order to render
understandable the physical analysis of the next section.

An {\it isodual field} $F^d=F^d(a^d,+^d,\times^d)$~\cite{ref2,ref10,ref11} is a
ring whose elements are the {\it isodual numbers}
\begin{equation}
a^d=a^{\dagger }\times 1^d=-a^\dagger,
\label{eq:two-1}\end{equation}
(where $a$ represents real numbers $n$, complex numbers $c$ or
quaternions $q$, $^\dagger$ represents Hermitean conjugation,
$1^d=-1$, and $\times$ represents the conventional multiplication) equipped
with: the {\it isodual sum}
\begin{equation}
a_1^d+^da_2^d=-(a_1^\dagger+a_2^\dagger),
\label{eq:two-2}\end{equation}
with {\it isodual additive unit} $0^d\equiv0$,
$a^d+^d0^d=0^d+^da^d\equiv a^d\;\forall a^d\in F^d$; the {\it isodual
multiplication}
\begin{equation}
a_1^d\times ^da^d_2=a_1^d\times (-1)\times a_2^d=-a_1^\dagger\times
a_2^\dagger,
\label{eq:two-3}\end{equation}
with {\it isodual multiplicative unit} (or {\it isodual unit} for short) $1^d=-1$,
$a^d\times ^d1^d=1^d\times ^d a^d\equiv a^d\;\forall a^d\in F^d$; and
remaining {\it isodual operations}, such as the {\it isodual quotient}
\[
a_3^d=a^d_1/^da_2^d=-a_1^d/a_2^d=-a_1^\dagger/a_2^\dagger,
\]
\begin{eqnarray}
& & a_3^d\times ^da_2^d=a_1^d,
\label{eq:two-4}\end{eqnarray}
the {\it isodual square root}
\[
(a^d)^{\frac{1}{2}d}=\sqrt{-a^d}=\sqrt{a^\dagger},
\]
\begin{equation}
(a^d)^{\frac{1}{2}d}\times ^d(a^d)^{\frac{1}{2}d}=a^d,
\label{eq:two-5}\end{equation}
and others~\cite{ref5,ref10,ref11}.

A property most
important for this note is that the {\it norm of isodual fields,
called isodual norm, is negative definite},
\begin{equation}
\mid a^d\mid ^d=\mid a^d\mid\times 1^d=-\mid a^d\mid=-(a\times
a^\dagger)^{\frac{1}{2}}.
\label{eq:two-6}\end{equation}

A quantity Q is called {\it isoselfdual} when it coincides with its isodual
\begin{equation}
Q \equiv Q^d = -Q^{\dagger}.
\label{eq:two-7}\end{equation}
For instance, the imaginary quantity $i=\sqrt{-1}$ is isoselfdual
because $i^d = -i^\dagger = -\overline{i} = -(-i) \equiv i$, where the upper
symbol $\overline{\  } $ denotes complex conjugation.

Note that for real numbers $n^d = -n$, and for complex number $c^d =
(n_1+i\times n_2)^d = n_1^d+^di^d \times ^dn_2^d = -n_1 + i\times n_2 =
-\overline{c}$. Note also that $\mid n^d \mid ^d = -\mid n \mid $ and
$\mid c^d \mid^d = -\mid \overline{c} \mid = -(n_1^2 +
n_2^2)^{\frac{1}{2}}$.

We finally note that {\it isodual fields satisfy all axioms of a field}
although in their isodual form. Thus, isodual fields
$F^d(a^d,+^{d},\times^d)$  are antiisomorphic to conventional fields
$F(a, +, \times)$, as desired.

For further studies on isodual fields the interested render may
consult \cite{ref5,ref10,ref11} (with the understanding that the
{\it isodual number theory} has not yet been investigated by mathematicians
until now).

An $n$--dimensional {\it isodual metric space} $S^d =
S^d(x^d,g^d,R^d)$ \cite{ref2,ref11} is a vector space with {\it isodual
coordinates} $x^d = -x = \{x^1,x^2,\ldots,x^n\}$ and {\it isodual metric}
$g^d=-g$ (where $g$ is an ordinary real and symmetric metric), defined
over the isodual real field $R^d = R^d(n^d,+^d,\times^d)$.

By recalling that the interval of a metric space must be an element of
the base field, the interval between two points $x_1^d, x_2^d \in S^d$
is given by
\begin{eqnarray}
(x_1^d-x_2^d)^{2d}& = &[(x_1^{id}-x_2^{id})\times g^d_{ij}
\times (x_1^{jd} - x_2^{jd})]\times 1^d = \nonumber \\
&=& [(-x_1^i+x_2^i)\times
(-g_{ij}) \times (-x_1^j+x_2^j)]\times (-1) \equiv \nonumber \\
 &\equiv &(x_1-x_2)^2.
\label{eq:two-8}\end{eqnarray}

We reach in this way the fundamental property of isodual theories
according to which the {\it interval of (real) metric spaces is
isoselfdual} (i.e., invariant under isoduality).

As important particular cases we have [loc.cit.]:
\begin{enumerate}
\item The 3--dimensional {\it isodual Euclidean space} $E^d =
E^d(r^d,\delta^d,R^d)$, $r^d = -r = -\{r^k\} = \{x^d,y^d,z^d\} =
\{-x,-y,-z\}$, $k = 1,2,3$, $\delta^d = -\delta = Diag(-1,-1,-1)$, with
{\it isodual sphere} $r^{d2d} = (-xx-yy-zz)\times (-I) \equiv r^2 =
(xx+yy+zz)\times (+I)$, with isodual unit $I^d = Diag(-1,-1,-1)$.
\item The $(3+1)$--dimensional {\it isodual Minkowski space} $M^d =
M^d(x^d,r^d,R^d)$, $x^d = -x = -\{x^{\mu}\} = -\{x^k,x^4\} = -\{r,c,t\}$,
where $c$ is the speed of light (in vacuum), and $\eta^d = -\eta =
-Diag(1,1,1,-1)$, with {\it isodual light cone} $x^{d2d} =
(x^{d\mu}\times\eta^d_{\mu\nu}\times x^{d\nu})\times I^d \equiv x^2$, $\mu,\nu
= 1,2,3,4$, with isodual unit $I^d = Diag(-1,-1,-1,-1)$.
\item The $(3+1)$--dimensional {\it isodual Riemannian space} $\Re ^d =
\Re ^d(x^d,g^d,R^d)$, $x^d = -x$ and $ g^d = -g(x)$ on $R^d$, where $g$
is a conventional Riemannian metric, with isodual interval $x^{d2d} =
(x^{d\mu}\ g^d_{\mu\nu}\ x^{d\nu})\times I^d \equiv x^2 = (x^\mu
g_{\mu\nu} x^\nu)\times(+I)$, with isodual unit $I^d =
Diag(-1,-1,-1,-1)$.
\end{enumerate}

The {\it isodual geometries} are the geometries of the isodual
spaces. This includes the
{\it isodual symplectic
geometry}~\cite{ref5,ref11}, which is the anti--isomorphic image of
the conventional symplectic geometry on the {\it isodual cotangent bundle}
$T^{d\ast} E^d(r^d,\delta^d,R^d)$ with 6--dimensional isodual unit $I^d_6
= I^d_3 \times I^d_3$.

The {\it isodual differential calculus} \cite{ref11} is characterized by
\begin{eqnarray}
d^{^d}x^k = -dx^k,& & d^{^d} x^{dk} = -d(-x^k) = dx^k \nonumber \\
  \partial^d f^d/^d\partial^dx^{dk}& = &-\partial f/\partial x^k.
\label{eq:two-9}\end{eqnarray}

 By recalling that p--forms must be elements of the base field, the
{\it  canonical one-form changes sign under isoduality},
\begin{equation}
\theta^d = (p_k^d\times^d d^dx^{dk})\times I^d_6 = -\theta ,
\label{eq:two-10}\end{equation}
while  the {\it canonical symplectic two-form is isoselfdual},
\begin{eqnarray}
\omega^d &=& (d^d x^{dk} \Lambda^d d^d p^d_k)\times I^d_6 \equiv
\nonumber \\
 & \equiv & (dx^k\Lambda dp_k)\times (+I_6) = \omega .
\label{eq:two-11}\end{eqnarray}
For  further details one may consult \cite{ref5,ref11}.

The {\it isodual Lie theory} \cite{ref5,ref12} is the anti-isomorphic image
of the conventional Lie theory under isoduality~\ref{eq:one-one}. The
{\it isodual enveloping association algebra} $\xi^{rd}$ is characterized by the
infinite-dimensional basis[loc.cit.]
\begin{equation}
\xi^d: I^d,\; \;  X^d_k, \; \; X_i^d\times ^dX_j^d, \; \; i\leq j, etc.
\label{eq:two-12}\end{equation}
where $i,j,k = 1,2,\ldots ,n, X^d = -X = -{X_k}, {X_k} $ is a
conventional (ordered) basis of an n-dimensional Lie algebra $L \approx
\xi^-$, and $I^d$ is the n-dimensional isodual unit, $I^d =
Diag(-1,$ $ -1, \ldots, -1)$.

The attached antisymmetric algebra is the {\it isodual Lie algebra}
$L^d\approx (\xi^d)^-$ with basis $X^d = -X$ and {\it isodual
commutators} [loc.cit.]
\begin{eqnarray}
L^d:& & [X_i^d,X_j^d]^d = X^d_i \times ^dX^d_j - X_j^d \times ^dX_i^d
= \nonumber \\
 & & =-(X_i) \times (-I) \times (-X_j) - (-X_j) \times (-I) \times
(X_i) =\nonumber \\
& & =C_{ij}^{dk} \times ^dX_k^d = -[X_i,X_j] = -C_{ij}^kX_k.
\label{eq:two-13}\end{eqnarray}

The {\it isodual exponentiation} is defined in terms of
basis~\ref{eq:two-12} [loc.cit.],
\begin{eqnarray}
\lefteqn{e^{d^{X^d}} = I^d+X^d/^d1!^d+X^d \times
^dX^d/^d2!^d+\ldots =}\nonumber \\
& & =(-I)(1+X/1!+X\times X/2!+\ldots ) = -e^{-X^d} = -e^X.
\label{eq:two-14}\end{eqnarray}

The (connected) {\it isodual Lie groups} $G^d$ [loc.cit.] as
characterized by the
isodual Lie algebra $L^d$ (under the conventional integrability
conditions of $L$ into $G$) are given by the isoexponential terms for
Hermitean generators $X=X^\dagger$
\begin{eqnarray}
G^d:& & U^d=\prod_k^d e^{\scriptstyle d^{\scriptstyle i\times^dw_k^d\times ^dX_k^d}} = \nonumber \\
& & = -\prod_k e^{\scriptstyle i\times w_k\times X_k} = -U,
\label{eq:two-15}\end{eqnarray}
where $w_k^k=-w_k \in R^d $ are the {\it isodual parameters} and  we have
used the isoselfduality of $i$.

It is evident that, for consistency, $G^d$ characterizes the {\it isodual
transforms} on $S^d(x^d,g^d,R^d)$
\begin{equation}
x^{d\prime} = U^d\times ^dx^d=-x^{\prime},
\label{eq:two-16}\end{equation}
 and that the {\it isodual group laws} are given by
\[
U^d(w_1^d)\times ^dU^d(w_2^d)=U^d(w_1^d+^dw_2^d),
\]
\begin{equation}
U^d(0^d)=I^d =-Diag(1,1,\ldots,1).
\label{eq:two-17}\end{equation}

For additional aspects of the isodual Lie theory one may
consult~\cite{ref5,ref12}.

An {\it isodual symmetry} is an invariance under an isodual group $G^d$. The
fundamental isodual symmetries are:
\begin{itemize}
\item[-] {\it isodual rotations} $0^d(3)$;
\item[-] {\it isodual
Euclidean symmetry} $E^d(3)=0^d(3)\times ^dT^d(3)$;
\item[-] {\it isodual Galilean
symmetry} $G^d(3.1)$;
\item[-] {\it isodual Lorentz symmetry} $L^d(3.1)$;
\item[-] {\it isodual
Poincar\'{e} symmetry} $P^d(3.1)=L^d(3.1)\times ^dT^d(3.1)$;
\item[-] {\it isodual
spin symmetry} $SU^d(2)$;
\item[-] {\it isodual spinorial Poincar\'{e} symmetry} ${\cal P}^d
=SL^d(2.C^d)\times ^dT^d(3.1)$;
\end{itemize}
and others [loc.cit].

The {\it isodual Hilbert space} ${\cal H}^d$ is characterized by the:
{\it isodual
states}
\begin{equation}
\mid \Psi>^d=-<\Psi\mid \; \; \; (or \Psi^d=-\Psi^\dagger);
\label{eq:two-18}\end{equation}
{\it isodual inner product}
\begin{equation}
<\Phi\mid \Psi>^d=<\Phi\mid ^d\times(I^{d-1})\times \mid \Psi>\times ^d I^d
=<\Psi \mid \times I^d \times \mid \Phi > \times I^d
\in
C^d(c^d,+,\times^d);
\label{eq:two-19}\end{equation}
and {\it isodual normalization }
\begin{equation}
<\Psi\mid \times (-I) \times \mid \Psi > = -I.
\label{eq:two-20}\end{equation}

The {\it isodual expectation values} of an operator $Q^d$ are given by
\begin{eqnarray}
<Q^d>^d&=&<\Psi\mid \times^dQ^d\times^d\mid\Psi>/^d<\Psi\mid \times(-I)
\times\mid\Psi>=
\nonumber \\
 &=& -<Q>,
\label{eq:two-21}\end{eqnarray}
where $<Q>$ is the conventional expectation value on a conventional
Hilbert space $\cal H$.

Similarly, the {\it isodual eigenvalue equations} for a Hermitean operator
$H=H^\dagger$ are given by
\begin{equation}
H^d \times^d\mid\Psi>^d=E^d\times^d\mid\Psi>^d,
\label{eq:two-22}\end{equation}
where $E^d=-E$.
One can therefore see that the isodual eigenvalues $E^d$ coincide with
the isodual expection values of $H^d$.

A property which is mathematically trivial, yet fundamental for the
physical analysis of this note is that the {\it normalization on a Hilbert
space is isoselfdual}~\cite{ref5},
\begin{eqnarray}
<\Psi\mid\Psi>^d & = & <\Psi\mid \times(-I)\times\mid\Psi>\times (-I)
\equiv
\nonumber \\
 &\equiv& <\Psi\mid \times(+I)\times\mid\Psi>\times (+I)=<\Psi\mid\Psi>.
\label{eq:two-23}\end{eqnarray}
The above property characterizes a new invariance which has remained
undetected since Hilbert's conception. Note, however, that, as it is
the case for the preceding novel invariance of the Minkowski line
element, the discovery of new laws~(\ref{eq:two-8}) and~(\ref{eq:two-23})
required the prior identification of {\it new numbers}, the isodual
numbers.

The theory of linear operators on a Hilbert space admits a simple, yet
significant isoduality~(see~\cite{ref5}) of which we can only mention
for brevity the {\it isodual unitary law}
\begin{equation}
U^d\times^dU^{\dagger d}=U^{t\dagger}\times^dU^d=I^d.
\label{eq:two-24}\end{equation}

Functional analysis also admits a simple, yet significant isoduality
which we cannot review for brevity [loc.cit.]. The non initiated reader  should
be alerted that, to avoid insidious inconsistencies, the {\it totality} of
conventional mathematical quantities, notions and operations must
be subjected to isoduality, e.g.: angles and related trigonometric
functions must be isodual; conventional and special functions and
transforms must be isodual; etc.~\cite{ref5}.

We finally recall that the isodual mathematics of this section admits
{\it three} sequential generalizations called {\it isotopic, genotopic} and
{\it hyperstructural} which we cannot review here for brevity~\cite{ref9,ref11}.

\section{ISODUAL THEORY OF ANTIMATTER }
The central assumptions of this note are~\cite{ref3}:
\begin{enumerate}
\item[1)]{\it matter} is represented by conventional mathematics, including
numbers, spaces, algebras, etc., based on the conventional positive
unit $+1$; while
\item[2)] {\it antimatter} is represented by the isodual mathematics of the
preceding section, including isodual numbers, isodual spaces, isodual
algebras, etc, based on the isodual unit $-1$.
\end{enumerate}

The above representations of matter and antimatter are then
interconnected by the isodual map~(\ref{eq:one-one}) which is
bi-injective and anti-isomorphic, as desired.

In this way, isoduality permits, apparently for the first time, a
representation of antimatter at all levels, beginning at {\it classical}
level and then continuing at levels of {\it first and second quantization}
in which it becomes equivalent to charge conjugation~\cite{ref4,ref9}.

By recalling that the isodual norm is negative definite,
Eq.~(\ref{eq:two-6}), an important consequence is that {\it all physical
characteristics which are positive for matter become negative for
antimatter}. The above occurrence is familiar with the charge $q$ whose
change  of sign in the transition from particles to antiparticles is
re-interpreted as isoduality, $q\rightarrow q^d=-q$. Jointly, however,
isoduality requires that the mass of antiparticles is negative,
$m^d=-m$, their energy is negative, $E^d=-E$, etc. Finally, isoduality
requires that antiparticles more backward in time, $t^d=-t$, as
originated conceived~\cite{ref1}.

One should note that the conventional positive values for particles
$m>0$, $E>0$, $t>0$, etc., are referred to corresponding positive
units, while the negative values for antiparticles, $m^d<0$, $E^d<0$,
$t^d<0$, etc., are referred to negative units of mass, energy, time,
etc. This implies the full equivalence of the two representations and
removes the traditional objections against negative physical
characteristics.

In fact, isoduality removes the historical reason that forced Dirac to
invent the ``hole theory''~\cite{ref1}, which subsequently restricted
the study of antimatter at the level of second quantization. We are
here referring to the fact that the negative energy solutions of
Dirac's equation behave unphysically when (tacitly) referred to
{\it positive units}, but they behave in a fully physical way when referred
to {\it negative units}~\cite{ref4,ref9}.

The isodual theory of antimatter {\it begins} at the primitive Newtonian
level so as to achieve a complete equivalence of treatments with
matter. The basic carrier space is the isodual space
\[
S^d(t^d,r^d,v^d)=  E^d(t^d,R^d)\times^dE^d(r^d,\delta^d,R^d)\times
^dE^d(v^d,\delta^d,R^d),
\]
\begin{equation}
V^d=d^dr^d/^dd^dt^d=-V,
\label{eq:three-1}\end{equation}
with corresponding total 7-dimensional (dimensionless) unit
\begin{eqnarray}
I^d&=&1^d_t\times^dI^d_r\times^dI^d_v, \nonumber \\
1^d_t=-1, I^d_r&=&-Diag(1,1,1)=I^d_v.
\label{eq:three-2}\end{eqnarray}
The fundamental dynamical equations are the {\it isodual Newton's equations}
first introduced by the author in~\cite{ref11}.
\begin{equation}
m^dx^dd^dv_k^d/^dd^dt^d=F^d_k(t^d,r^d,v^d), k=x,y,x,z.
\label{eq:three-3} \end{equation}

It is easy to see that the above theory represents correctly all
Newtonian Coulomb  interactions. In fact, the theory recovers the
{\it repulsive} Coulomb force between two charges $q_1$ and $q_2$ of equal
sign of matter, $F=k \times q_1\times q_2/r\times r >0$; it recovers
the {\it repulsive} force between the corresponding ``anti-charges'',
$F^d=k^d\times^dq_1^d\times^dq_2^d/^dr^d\times^dr^d<0$ (because now
the force is
referred to the unit -1); and recovers the {\it attractive} force between
charges $q$ and their conjugate $q^d$ when computed in our space,
$F=k\times q\times q^d/r\times r<0$ referred to the unit $+1$, or in
isodual space, $F^d=k^d\times^dq^d\times^dq/^dr^d\times^dr^d>0$ referred
to the unit $-1$.

Along similar lines, it is easy to see that the above theory recovers the
conventional Newtonian gravitational {\it attraction} for matter--matter
systems, $F^d=g\times m_1\times m_2/r\times r>0$; it predicts
gravitational {\it attraction} for antimatter--antimatter systems,
$F=g^d\times^dm_1^d\times^dm_2^d/^dr^d\times^dr^d<0$; and it predicts
gravitational {\it repulsion} (antigravity) for matter--antimatter systems,
$F^d=g\times m_1\times m_2^d/r\times r < 0 $ on $S(t, r, v)$ or
$F^d=g^d\times^dm_1^d\times^dm_2/^dr^d\times^dr^d>0$ on
$S^d(t^d,r^d,v^d)$.

The above predictions are confirmed at all subsequent classical
levels, including the representation of  antimatter on isodual
Riemannian spaces, which yields the characterization of antigravity
for matter--antimatter systems via the reversal of the sign of the
curvature tensor (see~\cite{ref5,ref9} for brevity). In different terms,
even thogh antimatter-antimatter systems are attractive as the ordinary
matter-matter ones, the gravitational fields of matter and of antimatter are
different. Thus, gravity can tell whether it is made up of matter or
antimatter.

The next level of study is that via the {\it isodual analytic
mechanics}~\cite{ref11}, which is characterized by the {\it isodual Lagrange
equations} (here omitted for brevity), and the {\it isodual Hamilton
equations} in the {\it isodual Hamiltonian}
$H^d(t^d,r^d,p^d)=-H(t,r,p)$~\cite{ref11}
\begin{equation}
\frac{d^dr^{kd}}{d^dt^d}d=\frac{\partial^dH^d}{\partial^dp_k^d}d,\:
\frac{d^dp_k^d}{d^dt^d}d=-\frac{\partial^dH^d}{\partial^dx^{kd}}d,
\label{eq:three-4}\end{equation}
which are defined on isodual space $S^d(t^d,r^d,p^d)$ with isodual
units~(\ref{eq:three-2}), an $p_k^d=m^d\times^dv^d_k$.

Eq.~(\ref{eq:three-4}) are derivable from the {\it isodual action }
\begin{equation}
A^d=\int_{\ }^d \; _{t_1^d}^{t_2^d} (p_k^d\times d^d x^
{kd}-H^d\times^dd^dt^d)=-A,
\label{eq:three-5}\end{equation}
where $\int ^d = -\int$ is the {\it isodual integral},
with {\it isodual Hamilton-Jacobi equations} [loc.cit.]
\[
\partial^dA^d/^d\partial^dt^d+H^d=0,
\]
\begin{equation}
\partial^dA^d/^d\partial^dr^{dk}-p_k^d=0.
\label{eq:three-6}\end{equation}

It is easy to see that the isodual analytic mechanics preserves all
electromagnetic and gravitational predictions of the isodual Newtonian
theory.

The operator formulation is characterized by a {\it new quantization for
antimatter}, which is missing in current theories. It can be first
expressed via the {\it naive isodual quantization}~\cite{ref11}
\begin{equation}
A^d\rightarrow i\times^d\hbar^d\times^d\ln ^d\Psi^d(t^d,r^d),
\label{eq:three-7}\end{equation}
under which Eq.~(\ref{eq:three-6}) are mapped into the {\it isodual
Schr\"{o}dinger equations}~\cite{ref11}
\begin{eqnarray}
\lefteqn{i\times^d\hbar^d\times^d\partial^d\Psi^d/^d\partial^dt^d=H^d\times
^d\Psi^d,}
\nonumber \\
& & p_k^d\times^d\Psi^d=-i\times^d\hbar^d\times^d\partial^d\Psi^d/^d\partial^dr^{dk},
\label{eq:three-8}\end{eqnarray}
with corresponding {\it isodual Heisenberg equation} [loc.cit.]
\begin{equation}
i\times^dd^dQ^d/^dd^dt^d=Q^d\times^dH^d-H^d\times^dQ^d.
\label{eq:three-9}\end{equation}

The above naive derivation is confirmed by the novel {\it isodual
symplectic quantization} which is not reviewed here for
brevity [loc.cit.].

Isodual techniques have therefore permitted the identification of a
hitherto unknown image of quantum mechanics called by the author
{\it isodual quantum mechanics}, whose structure is characterized
by~\cite{ref5}:
\begin{enumerate}
\item Isodual fields of real numbers $R^d(n^d,+^d,\times^d)$ and
complex numbers $C^d(c^d,+^d,\times^d)$.
\item Isodual carrier spaces, e.g., $E^d(r^d,\delta^d,R^d)$.
\item Isodual Hilbert space ${\cal H}^d$.
\item Isodual enveloping operator algebra $\xi^d$.
\item Isodual symmetries  realized via isodual unitary operators in
${\cal H}^d$, e.g., Eq.~(\ref{eq:three-1}).
\end{enumerate}

The fundamental notion of the theory is evidently given by the {\it isodual
Planck constant} $\hbar^d=-\hbar$, although referred to a negative unit
$-1$, thus being equivalent to $\hbar>0$ when referred to its positive
unit $+1$.

It is evident that the map from quantum mechanics to its isodual is
bi--injective and anti-isomorphic, as desired and as occurring at all
preceding levels.

Note in particular that the new Hilbert space invariance
law~(\ref{eq:two-23}) assures that all physical laws which hold for
particles also hold for antiparticles, as confirmed by the equivalence
between charge conjugation and isoduality.

It should be noted that charge conjugation is (bi-injective and)
{\it homomorphic} because spaces are mapped into themselves. On the
contrary, isoduality is (bi-injective and) {\it anti-isomorphic} because
spaces are mapped into new ones, the isodual spaces, which are
coexistent, yet physically distinct from conventional spaces. The
latter occurrence will soon appear crucial for the main results of this
note.

Intriguingly, {\it isodual quantum mechanics recovers the known
electromagnetic and weak phemenology of antiparticles}~\cite{ref5,ref9},
thus providing sufficient credibility for further studies. In
fact, the isodual operator theory merely provides a {\it re-interpretation}
of existing phenomenological knowledge, as the reader is encouraged to
verify, e.g., for the quantum Coulomb interactions.

It should be also noted that the above results are reached for
{\it antiparticles in first quantization}, because second quantization is
done for exactly the same reasons used for {\it particles}, no more and no
less, owing to the complete parallelism of the theories for matter and
antimatter at all levels.

Our relativistic theory of antimatter also begins at the classical
level and it is based on a new image of the special relativity called
by this author the {\it isodual special
relativity}~\cite{ref3,ref5,ref9}.
The latter theory is based on the isodual Minkowski space
$M^d(x^d,\eta^d,R^d)$ with basic isodual unit of space and time (also
in dimensional form) $I^d=-Diag(1,1,1,1,)$.

The fundamental symmetry is the isodual Poincar\'{e} symmetry
$P^d$(3.1)$\: =\: L^d$(3.1) $\times^d\: T^d$ (3.1), with {\it isodual Lorentz
transforms}
\begin{eqnarray}
\lefteqn{
\left\{ \begin{array}{rcl}
x'^{d1} & = & x^{d1}, \; \; \; x'^{d2} =  x^{d2}, \\
x'^{d3} & = & \gamma^d\times^d(x^{d3}-\beta^d\times^dx^{d4}), \\
x'^{d4} & = & \gamma^d\times^d(x^{d4}-\beta^d\times^dx^{d3}),
\end{array} \right. } \nonumber \\
& & \mbox{ } \; \beta^d  =  v^d/^dc^d=-\beta, \; \;
\beta^{d2d}=v^d\times^dv^d/^dc^d\times^dc^d=-\beta^2,
 \nonumber \\
& & \mbox{ } \; \gamma^d  =  1/^d(1-\beta^2)^{d\frac{1}{2}d}=-1/(1-\beta^2)^{1/2}=-\gamma,
\label{eq:three-10} \end{eqnarray}
where we have used properties~(\ref{eq:two-5}).

It is instructive to verify that transforms~(\ref{eq:three-10}) are the
negative version of the conventional transforms, thus confirming the
isodual Lie theory~\cite{ref12}.

The applicability of the isodual special relativity for the
characterization of antimatter is established by the isoselfduality of
the relativistic interval
\begin{eqnarray}
(x_1-x_2)^{d2d}& = &
[(x_1^{d\mu}-x_2^{d\mu})\times\eta^d_{\mu\nu}\times(x_1^{d\nu}-x_2^{d\nu})]\times
I^d\equiv \nonumber \\
 &\equiv & [(x_1^\mu-x_2^\mu)\times
\eta_{\mu\nu}(x_1^\nu-x_2^\nu)]\times I \nonumber \\
& = & (x_1-x_2)^2 ,
\label{eq:three-11}\end{eqnarray}
which has remained unknown throughout this century because of the
prior need of the isodual numbers.

It is an instructive exercise for the interested reader to see that
the isodual special relativity recovers all known classical
electromagnetic phenomenology for antiparticles~\cite{ref5,ref9}.

The isodual theory of antimatter sees its best expression at the level
of {\it isodual relativistic quantum mechanics}~\cite{ref5,ref9}, which is
given by a simple isoduality of the conventional theory here omitted
for brevity.

We merely point out that {\it negative units and related isodual theory
appear in the very structure of the conventional Dirac equation}
\[
{\gamma^\mu\times [p_\mu-e\times A_\mu(x)/c]+i\times
m}\times\Psi(x)=0,
\]
\begin{equation}
\gamma^k=\left( \begin{array}{cc}
0 & \sigma^k \\
-\sigma^k & 0
\end{array} \right) , \; \; \gamma^4=i\times\left(
\begin{array}{cc}
I_s & 0 \\
0 & -I_s \end{array} \right) .
\label{eq:three-12}\end{equation}
In fact, the {\it isodual unit of spin} $I^d=-I_s=-Diag(1,1)$ enters the
very structure of $\gamma^4$, while the {\it isodual Pauli matrices}
$\sigma^{kd}=-\sigma^k$ enter in the characterization of $\gamma^k$.

The above occurrence implies the emergence of a {\it novel interpretation} of
the {\it conventional} Dirac equation based on the following total unit,
space and symmetry
\begin{eqnarray}
I_{Tot}& = &\{I_{orb}\times I_{spin}\}\times
\{I^d_{orb}\times^dI^d_{spin}\}, \nonumber \\
M_{Tot}& = & \{M(x,\eta,R)\times
S_{spin}\}\times\{M^d(x^d,\eta^d,R^d)\times^dS^d_{spin}\} \nonumber \\
S_{Tot} & =& \{SL(2.C)\times T(3.1)\}\times
\{SL^d(2.C^d)\times^dT^d(3.1)\} \; \;,
\label{eq:three-13}\end{eqnarray}
where $I_{orb}=Diag(1,1,1,1)$ and $I_{spin}=Diag(1,1)$.

It should be indicated that the latter re-interpretation (which has
also escaped attention thoughout
 this century) is necessary for consistency. In fact, {\it the
conventional gamma matrices~(\ref{eq:three-12}) are isoselfdual}. The
conventional interpretation that the Poincar\'{e} symmetry $\cal P
$(3.1)$=SL$(2.C)$\times T$(3.1) is the symmetry of Dirac's equation
then leads to inconsistencies because $\cal P$(3.1) {\it is not}
isoselfdual. Only the product $\cal P$(3.1)$\times {\cal P}^d$(3.1) is
isoselfdual.

The above results permit the following novel re-interpretation of
Eq.s~(\ref{eq:three-12})
\[
{\tilde{\gamma}^\mu\times[p_\mu-e\times A(x)/c]+i\times
m}\times\tilde{\Psi}(x) = 0,
\]
\[
\tilde{\gamma}_k=\left( \begin{array}{cc}
0 & \sigma^d_k \\
\sigma_k & 0 \end{array} \right), \; \;
\tilde{\gamma}^4=i\left( \begin{array}{cc}
I_s^d & 0 \\
0 & I_s \end{array} \right),
\]
\begin{equation}
\{\tilde{\gamma}_\mu,\tilde{\gamma}_\nu\}=2\eta_{\mu\nu}, \; \;
\tilde{\Psi}=-\tilde{\gamma}_4\times\Psi=i\times
\left( \begin{array}{c}
\Phi \\ \Phi^d \end{array} \right)
\label{eq:three-14}\end{equation}
where $\Phi(x)$ is now two-dimensional.

Note that the above equations: remove the need of second quantization;
eliminate the need of charge conjugation because the antiparticle is
represented by $\sigma^d$, $I^d$, $\Phi^d$; and restore the correct
representation of spin $1/2$ via the conventional {\it two-dimensional}
regular representation of SU(2), rather than the current use
of a {\it four-dimensional} $\gamma$-representation.

We are now equipped to address the main objective of this note, a study
of {\it the anti-hydrogen atom and its spectroscopy}.

For this purpose we restrict our analysis to {\it massive particles} defined
as irreducible unitary representations of the Poincar\'{e} symmetry
$\cal P$(3.1). This essentially restricts the analysis to the electron
$e$ and the proton $p$, both considered as {\it elementary}, due to certain
ambiguities for composite hadrons indicated in App.A.

We then introduce the notion of {\it massive isodual particles} as
irreducible
unitary representations of the
isodual Poincar\'{e} symmetry ${\cal P}^d$(3.1). This again
restricts the analysis to the {\it isodual electron} $e^d$ and {\it isodual
proton} $p^d$.

At this point we should indicate the differences between
``antiparticles'' and ``isodual particles''. In first approximation,
these two notions can be identified owing to the equivalence of charge
conjugation and isoduality. However, at a deeper inspection,
antiparticles and isodual particles result to be different on a number
of grounds, such as:

\begin{enumerate}
\item Isoduality is broder than the PTC symmetry because, in
addition to reversing space-time coordinates $x\rightarrow x^d=-x$ (PT)
and conjugating the charge $q\rightarrow q^d=-q$ (C), it also reverses
the sign of the background units.
\item Antiparticles are defined in our own space-time, while isodual
particles are defined in a new space-time which is physically different
than our own.
\item Antiparticles have {\it positive} mass, energy, (magnitude of) spin,
etc., and {\it move forward in time}, while isodual particles have {\it
negative}
mass, energy, (magnitude of) spin, etc. and {\it move backward in time}.
\end{enumerate}

Next, we consider the bound states of the above elementary particles
which are given by: the hydrogen atom $A=(p, e)_{QM}$; the isodual
hydrogen atom $A^d=(p^d,e^d)_{IQM}$; and the positronium
$P=(e,e^d)_{QM}=(e^d,e)_{IQM}$, where QM (IQM) stands for Quantum
Mechanics (Isodual Quantum Mechanics).

It is evident that the hydrogen atom is characterized by the familiar
Schr\"{o}dinger's equation in the Coulomb spectrum $E_n$ on the
Hilbert space $\cal H$ ,
\begin{equation}
H\times \mid\Psi>=E_n\times\mid\Psi>.
\label{eq:three-15}\end{equation}

The isodual hydrogen atom is then characterized by the isodual image
on ${\cal H}^d$,
\begin{equation}
H^d\times^d\mid\Psi>^d=E^d_n\times^d \mid\Psi^d.
\label{eq:three-16}\end{equation}
The isodual theory therefore predicts that {\it the isodual hydrogen atom
has the same spectrum of the conventional atom, although with energy
levels reversed in sign}, $E^d_n=-E_n$.

{\it The positronium is an isoselfdual state}, because evidently invariant
under the interchanges $e\rightarrow e^d$, $e^d\rightarrow e$. As such,
it possesses a {\it positive} spectrum $E_n$ in our space-time and a
{\it negative} spectrum when studied in isodual space-time. In fact, the
total state of the positronium is given by $\mid Pos>\; =\;\mid e>\times \mid
e>^d$ with Schr\"{o}dinger's equation in our space-time $(\hbar=1)$
\begin{eqnarray}
\lefteqn{i\frac{\partial}{\partial t}\mid Pos>=(p_k\times
p^k/2m)\times\mid e>\times\mid e>^d+} \nonumber \\
 &  +\mid e>\times (p_k\times p^k/2m)^d\times^d\mid e>^d &+ \;
V(r)\times\mid e>\times\mid e>^d = \nonumber \\
 & &=E_n\mid Pos>, \; \; E_n>0,
\label{eq:three-17}\end{eqnarray}
with a conjugate expression in isodual space-time.

As indicated earlier, the isodual theory recovers the available
information on electromagnetic (and weak) interactions of
antiparticles. No novelty is therefore expected along these lines
in regard to the antihydrogen atom and the positronium.

However, the isodual theory has the following novel predictions for
gravitational interactions~\cite{ref6,ref9}:

\begin{quote}
PREDICTION I: {\it Massive stable isodual particles and their bound stated
(such as the isodual hydrogen atom) experience antigravity in the field
of matter and ordinary gravity in the field of antimatter.}
\end{quote}

\begin{quote}
PREDICTION II: {\it Bound states of massive stable particles and their
isoduals (such as
the positronium) experience ordinary gravity in
both fields of matter and antimatter.}
\end{quote}

We now remain with the central open problem raised in this note: {\it Does
antimatter emit a new light different than that emitted by ordinary
matter}?

The answer provided by the isodual theory is in the
affirmative. Recall that the {\it photon} $\gamma$ emitted by the hydrogen
atom has {\it positive} energy and time according to the familiar
plane-waves characterization on $M(x,\eta,R)$ with unit
$I=Diag(1,1,1,1)$
\begin{equation}
\Psi(t,r)=A\times e^{i\times(k\times r-E\times t)}.
\label{eq:three-18}\end{equation}

The isodual hydrogen atom has {\it negative} energy and time. As such, it
is predicted to
emit photons with the same characteristics, here called
{\it isodual photon} and characterized by the isoduality
\begin{eqnarray}
\Psi^d(t^d,r^d)&= & A^d\times^de^{\scriptstyle d ^{\scriptstyle
i^d\times^d(k^d\times^dr^d-E^d\times^d
t^d)}}= \nonumber
\\
 & = & A\times e^{\scriptstyle -i\times(k\times r^d-E\times t^d)},
\label{eq:three-19}\end{eqnarray}
now defined on $M^d(x^d,\eta^d,R^d)$ with isodual unit
$I^d=-Diag(1,1,1,1)$.

It is easy to see that the isodual photon is characterized by the
isodual special relativity via the spin $1^d=-1$, massless,
irreducible, isodual unitary representation of $P^d$(3.1), the {\it isodual
neutrinos} being the corresponding representation for spin
$\frac{1}{2}^d=-\frac{1}{2}$.

It should be recalled again that the isodual particle $\gamma^d$
behaves exactly like the ordinary photon $\gamma$ under all
electromagnetic and weak interactions. Moreover, massive particles and
their isoduals do annihilate into photons,
but photons and their isoduals cannot
annihilate.
The {\it sole} known possibility to distinguish photons from their isoduals
is via {\it gravitational} interactions. We reach in this way the following:

\begin{quote}
PREDICTION III: {\it The isodual photon experiences antigravity in the
field of matter  and gravity in the field of antimatter.}
\end{quote}

If confirmed by future theoretical and experimental studies, the above
prediction would permit to ascertain whether a far away galaxy or
quasar is made up of matter or antimatter by measuring whether its
light is attracted or repelled by the gravitational field of matter.

In principle, the above hypothesis can be experimentally tested in
contemporary astrophysics, provided that care is exercised in the
interpretation of the results. In fact, the gravitational {\it attraction} of
light emitted from a very distant galaxy by a closer galaxy
 {\it is not} evidence that they are
made up of matter, because both galaxies can be made-up of antimatter.

However, the astrophysical measure of gravitational {\it repulsion} of light
under the above conditions would be an experimental confirmation of our
isodual theory of antimatter, although we would not be in a position
of identifying which of the two galaxies is made-up of matter and
which is made up of antimatter.

The resolution of the latter issues will be possible when the
technology on the gravitational deflection of light permits the
measure of light deflected by fields we are sure to be made-up of
matter, such as a near-by planet or  the Sun.

It should be indicated that the deflection of light originating
from far away galaxies or quasars has indeed been measured and in all cases
it has shown to be attracted by matter. Therefore, no light repelled by
matter (or isodual light) has been detected until now to our best
knowledge.

In summary, the ``new physics of antimatter'' will remain unsettled
until we have final {\it experimental} results on the behavior in the
{\it gravitational} field of matter of: isodual photons, isodual electrons
(positions), isodual proton (antiproton), positronium, and isodual
(anti) hydrogen atom.

Note that the availability of several of the above experimental
resolutions, but not that for the positronium will still leave the
theory of antimatter essentially unsettled because of the several
conceivable alternatives which remain possible.

In closing we should point out that, by no means, all possible open
questions related to the {\it new} isodual theory of antimatter can be
resolved in this introductory {\it note}.

Among a virtually endless number of aspects to be studied, we mention the
need for the {\it isodual re-interpretation} of second quantization
which can be achieved via the separation of
retarded and advanced solutions as belonging to the Minkowski space and
its isodual, respectively, and the re-interpretation of the former
(latter) as characterizing particles (isodual particles).

Preliminary (unpublished) studies have indicated that the above
re-interpretation essentially leaves unchanged the numerical results for
electromagnetic and weak interactions all the way to the isodual
re-interpretation of Feynman diagrams, although this expectation must
evidently be confirmed by specific studies.

However, the re-interpretation implies significant changes, e.g., the
study whether a given photon is indeed such or it is an isodual
photon. As an example, the isodual interpretation of the two-photon
decay of the positronium requires the re-formulation
\begin{equation}
Pos.=(e,e^d)_{IQM}\rightarrow\gamma+\gamma^d.
\label{eq:three-20}\end{equation}

A similar result is expected for other decays, e.g., $\pi^0\rightarrow
\gamma+\gamma^d$ under the
condition that the $\pi^0$ is also a  bound state of a particle its
antiparticle.  We should recall that no conservation law of matter and,
independently, of antimatter, is possible because of their interconnection
via finite transition probabilities established beginning with Dirac's
equation. Therefore, the isodual theory also admits the three-body
decays of the positronium into $\gamma + \gamma + \gamma^d$ or
$\gamma + \gamma^d + \gamma^d$, e.g., when the total energy of the
two $\gamma$'s (of the two $\gamma^d$'s) of the three-body decay is equal
to the energy of the $\gamma$ (of the $\gamma^d$) of the two-body decay.

Along similar lines, under sufficient amounts of energy, particle-antiparticle
systems can also decay for the isodual theory into one single photon
or one single isodual photon. Similar situations permitted by the lack of
conservation of matter and, separately, of antimatter, also occur for the
decay of pions and in other interactions.

Note that {\it there is no new law of conservation of matter and,
separately, of antimatter}, because of the finite transition
probability among them established beginning with Dirac's equation. The
only possible conservation laws are the {\it conventional} ones is
our space
time and, separately,their isoduals for antimatter.

In addition to the isodual studies in second quantization, it is
necessary to conduct further studies on the {\it interior gravitational
problem} of matter and of antimatter via the isotopic
methods~\cite{ref5,ref9}.

In fact the latter studies have provided the strongest  avaible
evidence on the existence of antigravity originating from the
{\it identification} (rather than the ``unification'') of the (exterior)
gravitational field with the electromagnetic field originating the
mass considered~\cite{ref13}. It is evident that such identification implies
the equivalence of the two phenomenologies, that is, gravitation is
expected to have both attraction and repulsion in the same way as
occurring for electromagnetism.

The above argument is so forceful that the experimental establishment of
the lack of existence of antigravity may ultimately imply the need to
re-write particle physics, beginning with the electron, into a form in
which the mass does not possess an appreciable electromagnetic origin,
or it may imply the lack of physical validity of the mathematical
notion of isoduality.

Finally, additional studies are needed on the classical and quantum
isotopic representation of gravity and its isodual~\cite{ref9}, because these
studies contain all the preceding ones {\it plus} the inclusion of
gravitation which is embedded in the unit for matter and in the
isodual unit for antimatter.

Preliminary (unpublished) studies have indicated that the latter
approach confirm the results of this note, while permitting further
advances at the isooperator gravitational level, e.g., an
axiomatically consistent formulation of the PTC theorem inclusive of
gravitation.

We finally note that the possible lack of existence of antigravity for
the isudual photon {\it will not} invalidate the isodual theory,
because it will only imply the isoselfduality of the photon, that is,
presence in the photon of both retarded and advanced solutions, which
would remain separated for massive particles. Thus, antigravity may
exists for massive particles without necessarily existing for light.

Interested readers are encouraged to identify possible {\it
theoretical} arguments against antigravity for light emitted by
antimatter, with the clear understanding that the final scientific
resolution one way or the other can only be the {\it experimental}
one.

\vspace*{2cm}
\appendix
%{Appendix A:
{\Large \bf APPENDIX A:\\

 Gravitational problematic aspects of quark theories.}
\vspace*{1cm}

The ``new physics of antimatter'' is expected to have an impact on all
of elementary particle physics, because it focuses the attention on novel
gravitational, rather than familiar electroweak aspects.

An illustration is given by a necessary reformulation of contemporary
quark theories. In fact, gravitation is solely defined in our
space-time, while quarks are solely defined in the mathematical,
unitary, internal space, with no interconnection being possible due to
the O'Rafeirtaigh theorem.

It {\it necessarily} follows that {\it all particles made-up of quarks cannot
have any gravitation at all}, which is grossly contrary to
experimental evidence.

It should be indicated that O'Raifeartaigh's theorem has been superseded
by graded Lie algebras and related supersymmetries, in which case a connection
between space-time and internal symmetries is possible. However, the validity
of such interconnection would require the prior establishment of the physical
validity of supersymmetries and the existence of their predicted new
particles. Irrespective of that, a correct formulation of the gravity of
quarks within this latter setting is faced with serious technical problems
and it has not been achieved until now, to our best knowledge.

The above conclusion is confirmed by the well known fact that {\it
quarks
cannot be characterized by irreducible representations of the Poincar\'{e}
group}, that is, quark masses do not exist in our space-time, and are mere
parameters in unitary spaces.

Even assuming that the above fundamental problem is somewhat resolved
via hitherto unknown manipulations, additional equally fundamental
problems exist in the construction of a quark theory of antimatter,
because it does not yield in general an anti-isomorphic image of the
phenomenology of matter.

When including the additional, well known problematic aspects of quark
theories (e.g., the vexing problem of confinement which is not
permitted by the uncertainty principle), a structural revision of
contemporary quark theories becomes beyond {\it credible} doubts.

The only resolution of the current scientific impass known to this
author is that advocated since 1981, {\it quarks cannot be elementary
particles}~\cite{ref14}, as apparently confirmed by recent experiments at
Fermi-lab~\cite{ref15}.

In fact, the compositeness of quarks would permit their construction
as suitable bound states of physical massive particles existing in our
space-time, in which case (only) there would be the regaining of the
physical behavior under gravity.

The following aspects should however be clearly stated to separate
science from fiction. First, the above new generation of quark
theories requires the abandonment of the conventional Poincar\'{e}
symmetry $P$(3.1) in favor of a nonlinear, nonlocal-integral and non canonical
generalization, e.g., the isopoincar\'{e}-symmetry
$\widehat{P}$(3.1)$\approx P$(3.1)~\cite{ref16}. In
fact, a consistent construction of composite quarks inside hadron
requires the necessary alteration of the {\it intrinsic}
characteristics of ordinary
particles which is prohibited by $P$(3.1) but rather natural for
$\widehat{P}$ (3.1)~\cite{ref16} and related methodology~\cite{ref5}.

The use of the $q-$, $k-$ and quantum deformations of should
be excluded because afflicted by excessive problems of physical
consistency (which are absent for isotopies), such as~\cite{ref5}:
\begin{itemize}
\item[1)] Lack of invariance of the basic unit with consequential
inapplicability to actual measurements;
\item[2)] Lack of preservation of Hermiticity in time with
consequential lack of observables;
\item[3)] Lack  of invariant special functions (because, e.g., the
number $q$ becomes an operator under the time evolution).
\item[4)] Lack of uniqueness and invariance of physical laws;
\item[5)] Loss of Eistein's axioms; etc.
\end{itemize}

Second, the real constituents of hadrons are expected to be the quark
{\it constituents}
 and not the quarks themselves~\cite{ref14}. This new perspective
removes altogether the need for confinement. As a matter of fact, the
hadronic constituents are expected to be produced free and actually
identified in the massive particles produced in the spontaneous decays
with the lowest mode~\cite{ref16}.

The latter particles become conceivable as constituents because of the novel
renormalizations of their {\it intrinsic} characteristics which are
permitted by internal nonlagrangian and nonhamiltonian effects.

Third, the primary physical meaning of unitary theories and related
methodologies (Poincar\'{e} symmetry, SU(3) symmetry, relativistic quantum
mechanics, etc.) is their historical one: having achieved the final
classification of hadrons into families and the final understanding of
the related exterior phenomenology.

In much of the way as it occurred for the atoms in the transition from
the Mendeleev {\it classification} into families to the different problem of
the {\it structure} of the individual atoms, the transition from the unitary
classification of hadrons into families to the different problem of
the structure of the individual hadrons, is expected to require a
nonlinear, nonlocal-integral and nonpotential-nonhamiltonian
generalization of relativistic quantum mechanics, e.g., of the
isotopic axion-preserving type of ref.s~\cite{ref5,ref14,ref16}.

We should not forget that hadrons are not ideal spheres with points in
them, but are instead some of the densest objects measured in
laboratory by mankind in which the constituence are in a state of
total mutual penetration of the wavepackets. It is an easy prediction
that, even though of clear preliminary physical value, the use for the
latter conditions of theories which
are linear, local-differential and Lagrangian-Hamiltonian will not
resist the test of time.

The author would be grateful to colleagues who care to bring to his
attention any credible alternative to the above lines~\cite{ref5,ref14},
that is, a new
generation of theories with composite quarks which:
\begin{enumerate}
\item[1)] admit physical
constituents unambiguously defined in our space-time;
\item[2)] represent without
ambiguities the gravitational behavior of matter and antimatter; and
\item[3)] are based on the {\it exact} validity of the
 Poincar\'{e}-symmetry, quantum
mechanics and all that.
\end{enumerate}

\vspace*{2cm}
\centerline{\bf \large Acknowledgments}

\vspace*{0.5cm}

The author would like to thank for critical comments all
participants to the ``International Workshop on Antimatter Gravity and
Antihydrogen Atom Spectroscopy'', Sepino (IS), Italy, May
1996. Particular thanks are due to Allen P. Mills, jr., for invaluable
critical comments. Special thanks are finally due to the referee for a very
accurate reading of the manuscript, for spotting a number of misprints in
origin version and for the sound request of several clarifications which have
 permitted a clear improvement of the presentation.


\begin{thebibliography}{00}
\bibitem{ref1}
P.A.M.Dirac, The Principles of Quantum Mechanics, (Clarendon Press,
Oxford, 1958);
\\
R.L.Forward in Antiproton Science and Technology, ed. B.W.Augenstein,
B.E.Bonner, F.E.Mill and M.M.Nieto (Word Scientific, Singapore, 1988).
\bibitem{ref2}
R.M.Santilli, Hadronic J. 8 (1985), 25 and 36.
\bibitem{ref3}
R.M.Santilli, Isotopic Generalization of Galileis and Einstein's
Relativities, vol-s I and II (Hadronic Press, Palm Harbor, FL, 1991).
\bibitem{ref4}
R.M.Santilli, Comm.Theor.Phys. 3 (1994), 153.
\bibitem{ref5}
R.M.Santilli, Elements of Hadronic Mechanics, vol-s I (1994), II (1995),
III(in preparation)(Hadronic Press, Palm Harbor, FL).
\bibitem{ref6}
R.M.Santilli, Hadronic J. 17 (1994), 257.
\bibitem{ref7}
A.P.Mills, Hadronic J. 19 (1996), 79.
\bibitem{ref8}
M.Holzscheither, Hyp.Int., in press.
\bibitem{ref9}
R.M.Santilli in New Frontiers in Hadronic Mechanics,
ed. T.Gill(Hadronic Press, Palm Harbor, FL, 1996).
\bibitem{ref10}
R.M.Santilli, Algebras, Groups and Geometries 10 (1993), 273.
\bibitem{ref11}
R.M.Santilli, Rendiconti Circolo Matematico Palermo,
Suppl. 42 (1996),7.
\bibitem{ref12}
J.V.Kadeisvili, Rendiconti Circolo Matematico Palermo,
Suppl. 42 (1996),83.
\bibitem{ref13}
R.M.Santilli, Ann.Phys.(M.I.T.) 83 (1974), 108.
\bibitem{ref14}
R.M.Santilli, Found.Phys. 11 (1981), 383; Comm.Theor.Phys. 4 (1995),
123.
\bibitem{ref15}
Fermilab report no. 735 (1996).
\bibitem{ref16}
R.M.Santilli, Nuovo Cimento Lettere 37 (1983), 545; J. Moscow
Phys. Soc. 3 (1993), 255; Chinese J.Syst.Eng.and Electr. 6 (1995), 177.
\end{thebibliography}
\end{document}